\documentclass[usenatbib]{mn2e}
\usepackage{graphicx}

\title[Exoplanet Eccentricity Distribution]{The Exoplanet Eccentricity
  Distribution from Kepler Planet Candidates}

\author[Stephen R. Kane et al.]{Stephen R. Kane, David R. Ciardi, Dawn
  M. Gelino, Kaspar von Braun\\
  NASA Exoplanet Science Institute, Caltech, MS 100-22, 770
  South Wilson Avenue, Pasadena, CA 91125, USA}

\begin{document}

\maketitle

%%%%%%%%%%%%%%%%%%%%%%%%%%%%%%%%%%%%%%%%%%%%%%%%%%%%%%%%%%%%%%%%%%%%

\begin{abstract}

The eccentricity distribution of exoplanets is known from radial
velocity surveys to be divergent from circular orbits beyond
0.1~AU. This is particularly the case for large planets where the
radial velocity technique is most sensitive. The eccentricity of
planetary orbits can have a large effect on the transit probability
and subsequently the planet yield of transit surveys. The Kepler
mission is the first transit survey that probes deep enough into
period-space to allow this effect to be seen via the variation in
transit durations. We use the Kepler planet candidates to show that
the eccentricity distribution is consistent with that found from
radial velocity surveys to a high degree of confidence. We further
show that the mean eccentricity of the Kepler candidates decreases
with decreasing planet size indicating that smaller planets are
preferentially found in low-eccentricity orbits.

\end{abstract}


%%%%%%%%%%%%%%%%%%%%%%%%%%%%%%%%%%%%%%%%%%%%%%%%%%%%%%%%%%%%%%%%%%%%

\section{Introduction}
\label{intro}

Planets discovered using the radial velocity (RV) method have
dominated the total exoplanet count until recently, when the transit
method has made increasing contributions. The long time baseline of RV
surveys has allowed the detection more diverse orbital geometries than
achievable by ground-based transit surveys. The Kepler mission,
however, with its multi-year baseline, can begin to probe into
parameter space previously reserved for RV studies. At longer periods,
orbits tend to diverge significantly from the circular case beyond a
semi-major axis of $\sim 0.1$~AU \citep{but06}, although there may be
small observational biases that skew this distribution
\citep{she08}. This insight has led to numerous attempts to account
for eccentricity in the context of planet formation and orbital
stability \citep{for08b,mal08,mat08,tre08,wan11} and the influence of
tidal circularization \citep{pon11}.

It has been shown how eccentricity distribution effects transit
probabilities \citep{kan08,kan09} and projected yields of transit
surveys \citep{bar07,bur08}. This influence is minor for the
ground-based surveys since they are primarily sensitive to giant
planets in short-period orbits. However, the Kepler mission is
expected to be impacted by this distribution since it probes out to
much longer periods with a much reduced disadvantage of a window
function that affects observations from the ground \citep{von09}. A
comparison of the Kepler results in the context of eccentricity and
transit durations with the RV distribution has been suggested by
\citet{for08a} and \citet{zak11} and carried out by \citet{moo11}, but
initial planet candidate releases by the Kepler project do not provide
enough period sensitivity \citep{bor11a,bor11b}. The most recent
release of Kepler planet candidates by \citet{bat12} increases the
total number of candidates to more than 2,300 and the time baseline
probed to beyond 560 days. This has several implications for studies
of eccentricity distributions. The Kepler mission is sensitive to
planets significantly smaller than those accessible by current RV
experiments and thus allows a more in-depth study of the dependence of
eccentricity on the planet mass/size and multiplicity. If the
eccentricity distributions of Kepler and RV planets were found to be
substantially different then this may reveal a selection effect in the
way Kepler candidates are selected which is biased against eccentric
orbits. A direct comparison of the two distributions, provided they
are consistent for the planet mass/size region where their
sensitivities overlap, will allow a more exhaustive investigation of
orbital eccentricity to be undertaken.

Here we present a study of the eccentricity distribution of planets
discovered with the RV method and the complete list of Kepler planet
candidates. We calculate expected transit durations for circular
orbits and compare them with either calculated or measured eccentric
transit durations (\S \ref{eqns}). Our results show that the measured
transit durations from RV data (\S \ref{rv}) and the Kepler candidates
(\S \ref{kepler}) are consistent with having the same distribution. We
estimate the impact parameter distribution for the Kepler candidates
and show that their mean eccentricity decreases with decreasing planet
size (\S \ref{correlation}), which supports the hypothesis that
smaller planets tend to be found in multiple systems in near-circular
orbits. We discuss additional astrophysical aspects in \S
\ref{discussion} and conclude in \S \ref{conclusion}.

%%%%%%%%%%%%%%%%%%%%%%%%%%%%%%%%%%%%%%%%%%%%%%%%%%%%%%%%%%%%%%%%%%%%

\section{Eccentricity and Transit Duration}
\label{eqns}

A concise description of exoplanetary transit modeling and associated
parameters is presented elsewhere \citep{man02,sea03}. Here we
concentrate on the relevant details to our analysis: transit duration
and eccentricity. The transit duration for a circular orbit is given
by
\begin{equation}
  t_{\mathrm{circ}} = \frac{P}{\pi} \arcsin \left(
  \frac{\sqrt{(R_\star + R_p)^2 - a^2 \cos^2 i}}{a} \right),
  \label{duration}
\end{equation}
where $P$ is the orbital period, $a$ is the semi-major axis, $i$ is
the orbital inclination, and $R_\star$ and $R_p$ are the stellar and
planetary radii respectively. The impact parameter of a transit is
given by
\begin{equation}
  b \equiv \frac{a}{R_\star} \cos i
  \label{impact}
\end{equation}
and is defined as the projected separation of the planet and star
centers at the point of mid-transit.

\begin{figure*}
  \begin{center}
    \begin{tabular}{cc}
      \includegraphics[angle=270,width=8.2cm]{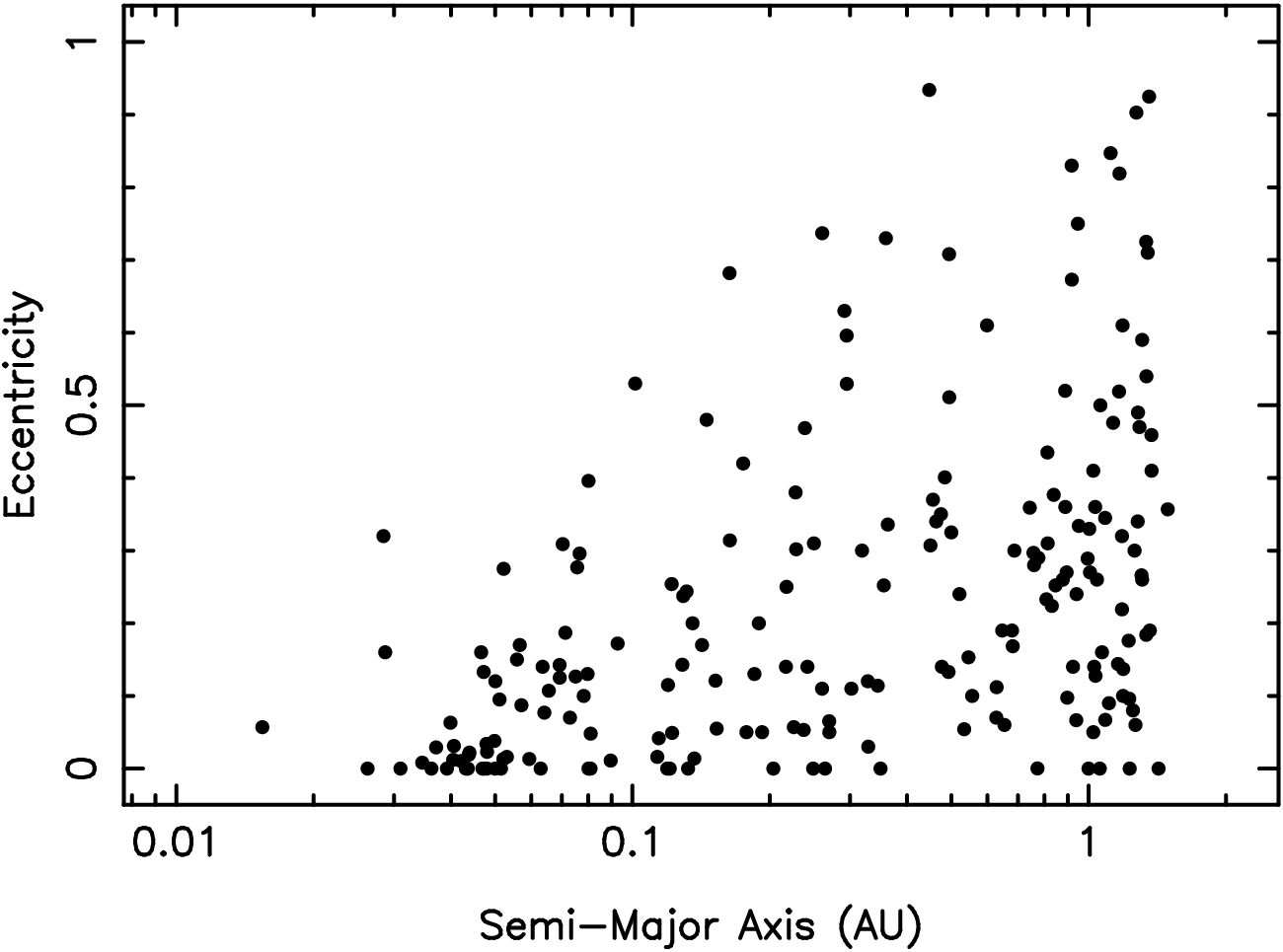} &
      \includegraphics[angle=270,width=8.2cm]{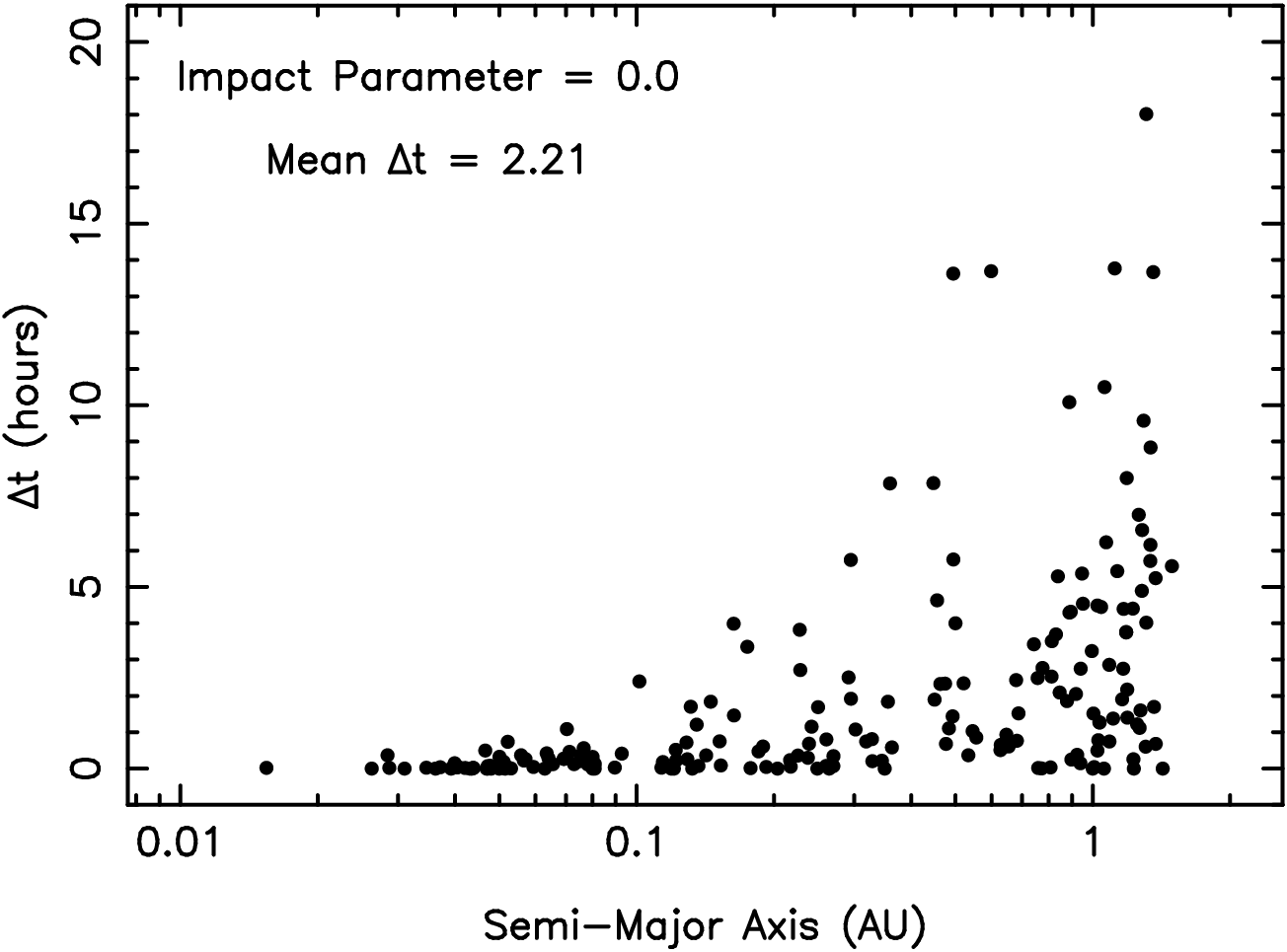} \\
      \includegraphics[angle=270,width=8.2cm]{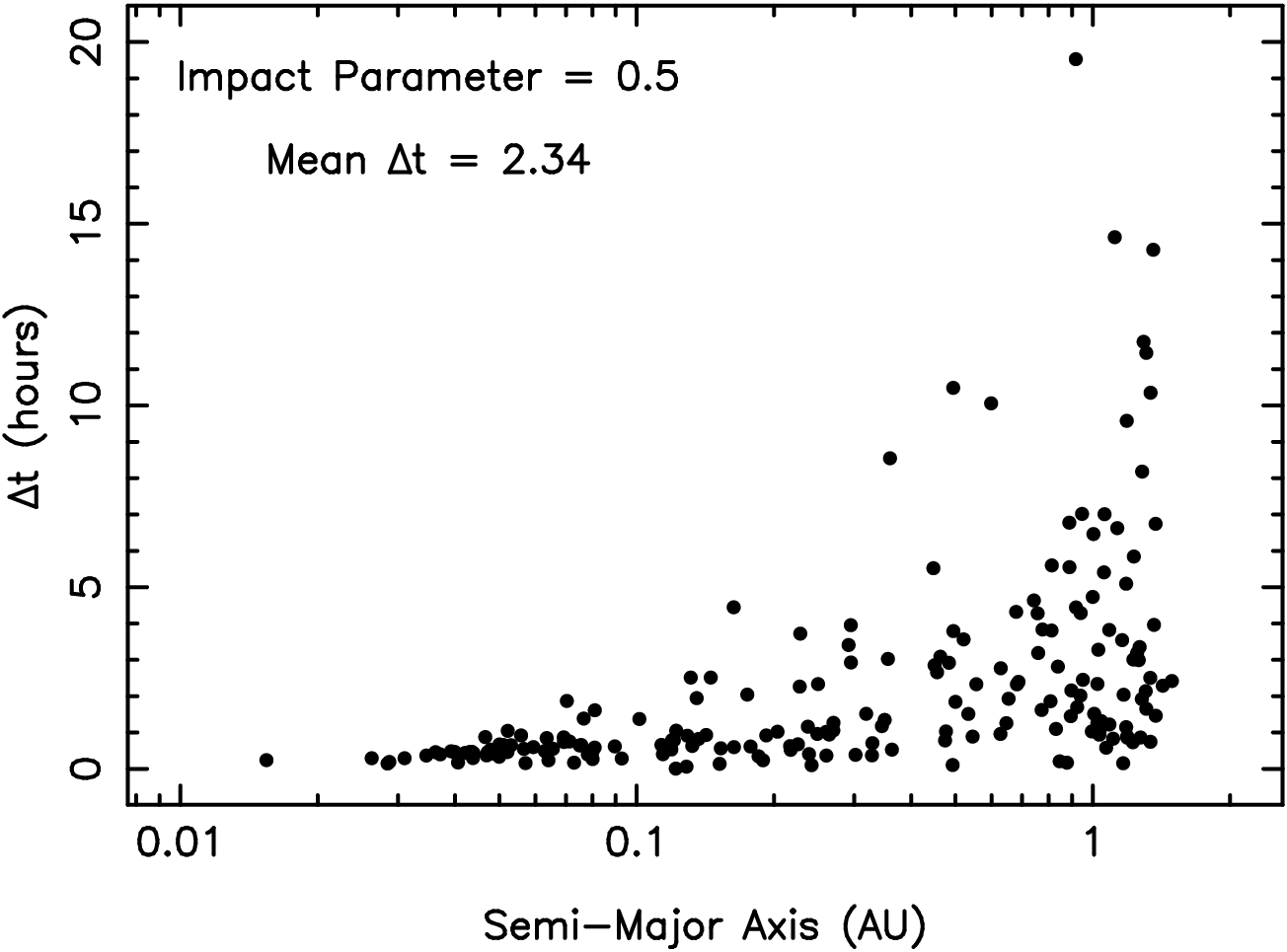} &
      \includegraphics[angle=270,width=8.2cm]{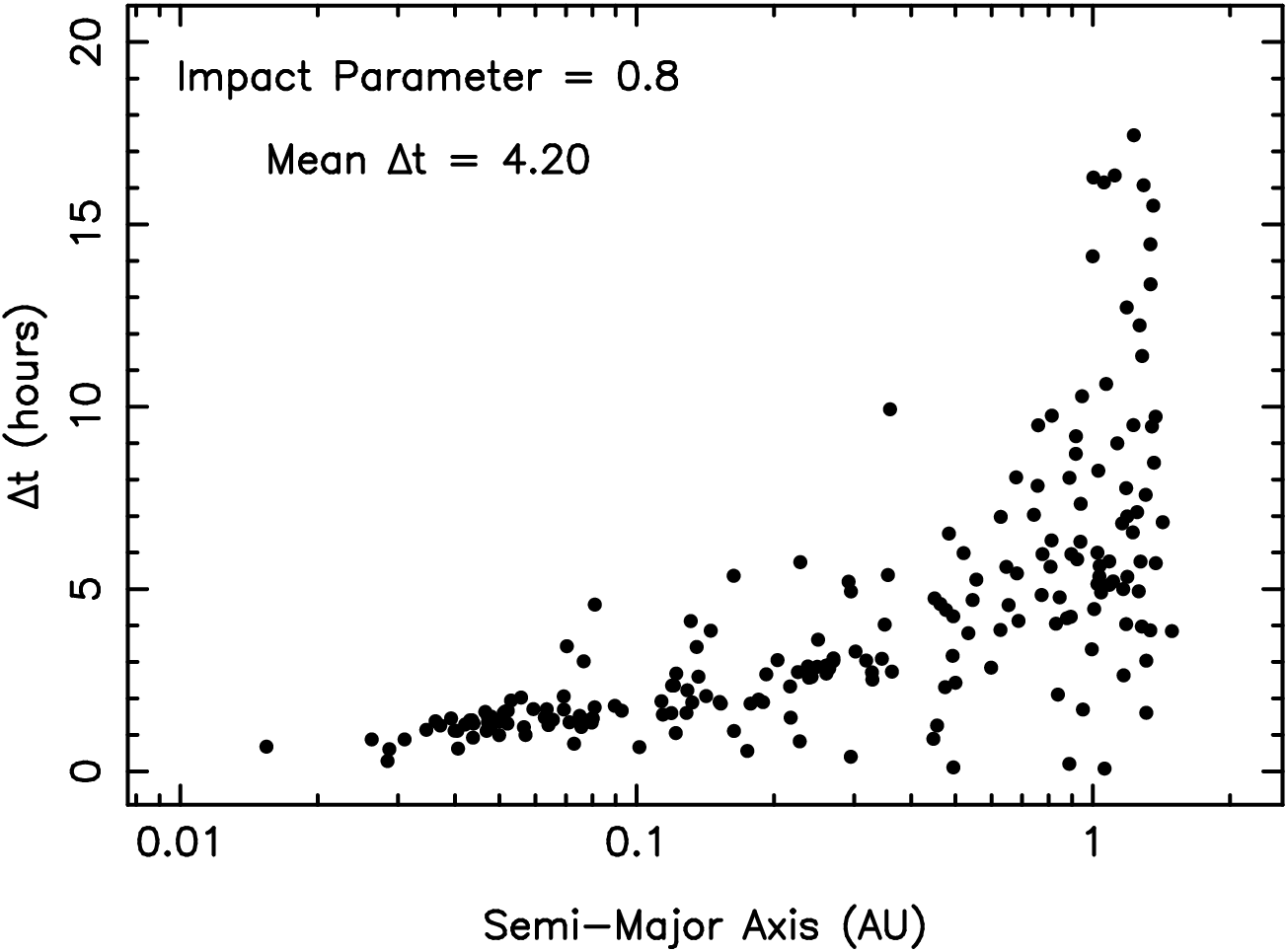} \\
    \end{tabular}
  \end{center}
  \caption{The eccentricity distribution of the known radial velocity
    planets (top-left) and the calculated transit duration difference
    (circular vs eccentric) for $b=0.0$ (top-right), 0.5
    (bottom-left), and 0.8 (bottom-right). See \S \ref{rv} for
    details.}
  \label{rvplanets}
\end{figure*}

For an eccentric orbit, the star--planet separation $r$ is
time-dependent and is given by
\begin{equation}
  r = \frac{a (1 - e^2)}{1 + e \cos f}
  \label{separation}
\end{equation}
where $e$ is the orbital eccentricity and $f$ is the true
anomaly. Replacing $a$ with $r$ at the time of inferior conjunction in
equations \ref{duration} and \ref{impact} provides generalized
expressions for transit duration and impact parameter for non-circular
orbits. \citet{bur08} converts these expressions into the scaling
factor
\begin{equation}
  \frac{t_{\mathrm{ecc}}}{t_{\mathrm{circ}}} = \frac{\sqrt{1-e^2}}{1 +
    e \cos(\omega-90\degr)},
    \label{scaling}
\end{equation}
where $\omega$ is the periastron argument of the orbit.

%%%%%%%%%%%%%%%%%%%%%%%%%%%%%%%%%%%%%%%%%%%%%%%%%%%%%%%%%%%%%%%%%%%%

\section{Radial Velocity Eccentricity Distribution}
\label{rv}

We first investigate the eccentricity distribution of the planets
discovered with the RV technique and the subsequent impact on the
predicted transit duration. The Exoplanet Data Explorer
(EDE)\footnote{\tt http://exoplanets.org/} stores information only for
those planets that have complete orbital solutions and thus are well
suited to this study \citep{wri11}. The EDE data are current as of
2012 February 24 and include 204 planets after the following criteria
are applied: $\log g > 3.5$ to exclude giant stars and $a < 1.5$~AU to
produce a sample that covers the same region in parameter space as the
Kepler candidates.

To calculate the transit duration, one needs an estimate of the
planetary radius. For planets that are not known to transit, we
approximate the planetary radius using the simple model described by
\citet{kan12}. This model adopts a radius of 1 Jupiter radius for
masses $ \geq 0.3 M_{Jupiter}$ and utilizes a power law fit to the
masses and radii of the known transiting planets for masses $< 0.3
M_{Jupiter}$. In order to estimate the radius of the host star, we use
the following relation related to the surface gravity
\begin{equation}
  \log g = \log \left( \frac{M_\star}{M_\odot} \right) - 2 \log
  \left( \frac{R_\star}{R_\odot} \right) + \log g_\odot
\end{equation}
where $\log g_\odot = 4.4374$ \citep{sma05}. Using the equations of
Section \ref{eqns} and the measured orbital parameters, we calculate
the transit duration for both the circular and eccentric cases. We
then take the absolute value of the difference between the two
durations as a diagnostic for the eccentricity distribution.

\begin{figure*}
  \begin{center}
    \begin{tabular}{cc}
      \includegraphics[angle=270,width=8.2cm]{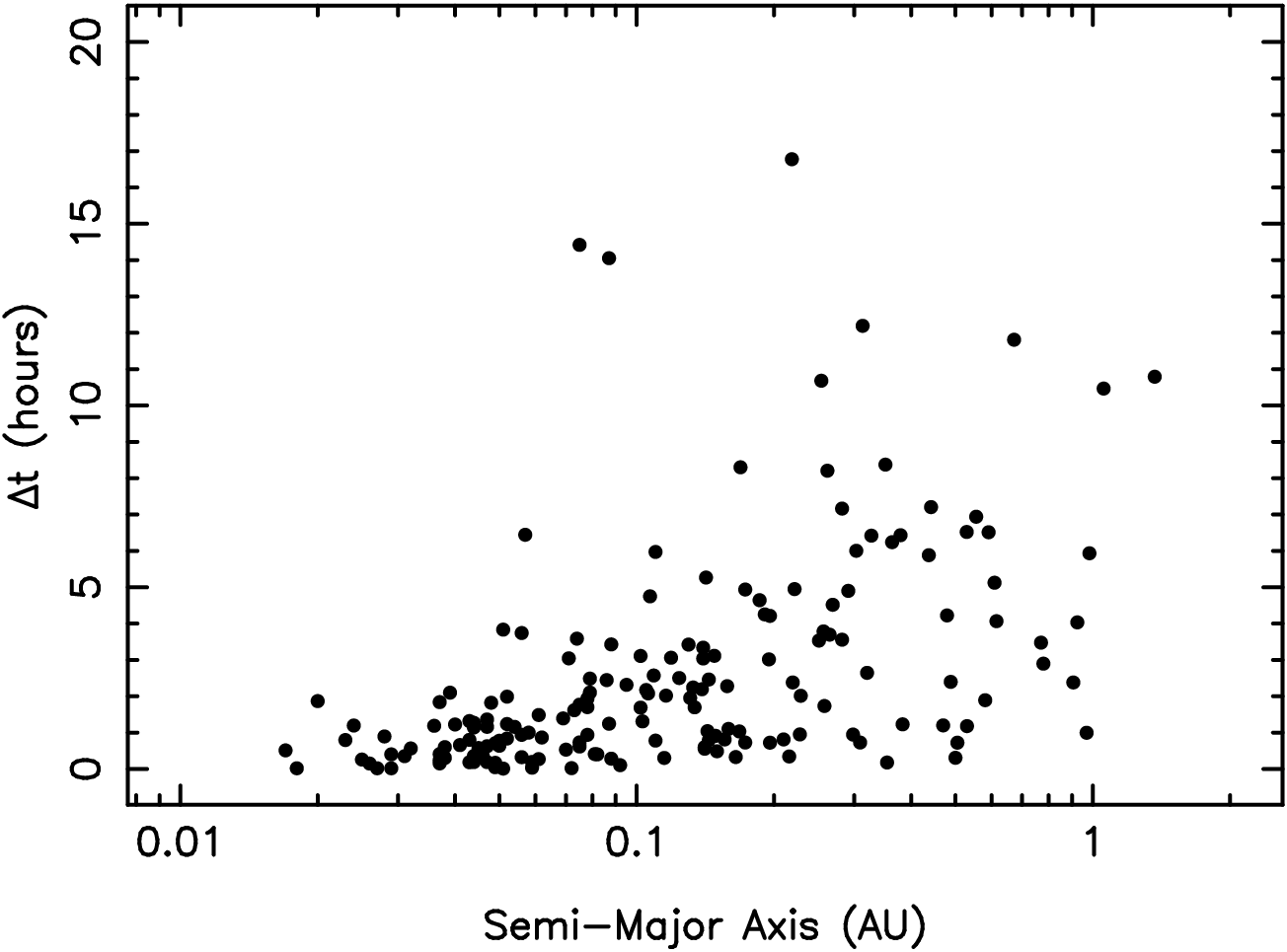} &
      \includegraphics[angle=270,width=8.2cm]{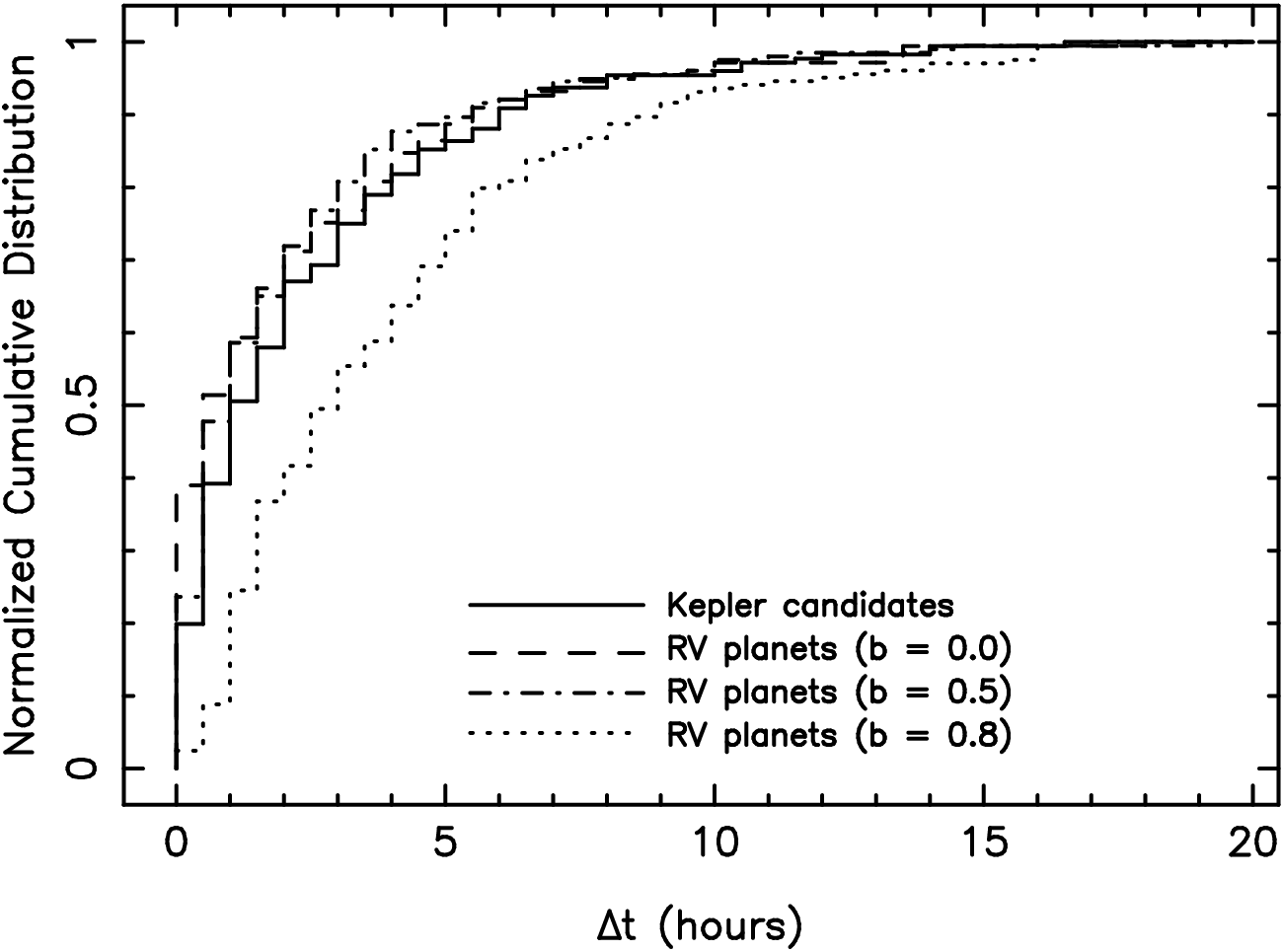}
    \end{tabular}
  \end{center}
  \caption{The calculated transit duration difference (circular vs
    measured) for the Kepler candidates (left). The cumulative
    histograms for the Kepler candidates and the three radial velocity
    planets shown in Figure \ref{rvplanets} show a close match
    between the distributions, quantified by the K-S test described in
    Section \ref{kepler}.}
  \label{keplercand}
\end{figure*}

The top-left panel of Figure \ref{rvplanets} shows the eccentricity
distribution of the RV planets taken from EDE as a function of
$a$. The distribution begins to diverge from mostly circular orbits
beyond 0.04~AU, and by 0.1~AU it has an eccentricity range of
0.0--0.5. This distribution is widely attributed to tidal damping of
the orbits after the disk has dissipated. According to \citet{gol66},
the timescale for orbital circularization is $\propto a^{6.5}
M_\star^{-1.5}$ where $M_\star$ is the stellar mass. Note that the two
planets inside of 0.04~AU with $e > 0.1$ are GJ~436b and GJ~581e, both
of which have M dwarf host stars.

The other three panels in Figure \ref{rvplanets} show the calculated
duration difference by $\Delta t$ as a function of $a$. $\Delta t$ is
the absolute value of the difference between the calculated transit
duration for a circular orbit and the calculated transit duration
based upon the measured orbital parameters (i.e., $\Delta t = |
t_{\mathrm{circ}} - t_{\mathrm{ecc}} |$), which is indicative of the
divergence from the assumption of only circular orbits. The top-right
panel assumes edge-on orbits ($i = 90\degr; b = 0$) for both the
circular and eccentric cases.  We show the effect of increasing the
impact parameter of the transits for $b = 0.5$ and $b = 0.8$. The mean
of $\Delta t$ is not significantly changed except for relatively high
values of $b$. We evaluate the significance of this distribution in
the following section.

It should be noted that, in order for a transit to take place, $i$ can
at most only slightly deviate from $90\degr$.  The consequent small
angle approximation means that the uniform distribution of $i$ values
maps to a uniform distribution of $b$ values, making all values of $b$
equally likely to occur.

%%%%%%%%%%%%%%%%%%%%%%%%%%%%%%%%%%%%%%%%%%%%%%%%%%%%%%%%%%%%%%%%%%%%

\section{Analysis of Kepler Transit Durations}
\label{kepler}

The release of more than 2,300 Kepler candidates is described in
detail by \citet{bat12}. The appendix table that contains the
characteristics of the Kepler candidates was extracted from the NASA
Exoplanet Archive\footnote{\tt
  http://exoplanetarchive.ipac.caltech.edu/}. We perform a similar
calculation for $\Delta t$ as described in the previous
section. However, this time we calculate the difference between
$t_{\mathrm{circ}}$ and the duration measurement provided by the
candidates table, $t_{\mathrm{kepler}}$. We do not use the provided
$b$ values since they are based on a circular orbit assumption from
the measured transit duration and the stellar radii. We thus make no
assumption on the value of $b$. We require $R_\star > 0.7 R_\odot$ to
remove candidates for which the uncertainty in the stellar radius
determination contributes significantly to the uncertainty in the
transit duration. We also only include candidates for which $R_p > 8
R_\oplus$ to limit the sample to giant planets, similar to the RV
sample. We show $\Delta t$ versus $a$ for the 176 resulting
candidates in the left panel of Figure \ref{keplercand}.

To compare this distribution to its equivalents in Section \ref{rv} we
perform a null hypothesis Kolmogorov-Smirnov (K-S) test to assess the
statistical significance of their similarities. We binned the data by
$\Delta t$ into 40 equal bins of 0.5 to collapse the data into
1-dimensional samples. The right panel of Figure \ref{keplercand}
shows the normalized cumulative histograms for each of the four
distributions: the three K-S tests compare the Kepler candidates to
the RV planets with assumed $b = 0.0$ (Test 1), $b = 0.5$ (Test 2),
and $b = 0.8$ (Test 3). Test 1 produces a K-S statistic of $D = 0.05$
that indicates a 100\% probability that these data are consistent with
being drawn from the same distribution (the null hypothesis). Test 2
produces a similar result of $D = 0.075$, also a probability of
100\%. Test 3 results in $D = 0.2$ which is equivalent to a
probability of 36\%. This result can readily be seen in the right
panel of Figure \ref{keplercand} where the $b = 0.0$ and $b = 0.5$
cases are almost indistinguishable from the Kepler candidates, but the
$b = 0.8$ case is clearly discrepant. As mentioned in \S \ref{rv}, the
small range of values for $i$ for transits results in a uniform
distribution of impact parameters with a mean of $b = 0.5$. The
statistical congruence in the K-S test implies that the Kepler Mission
is indeed recovering the eccentricity distribution of the RV planets.

A criticism that may be levelled at this methodology is that the
outcome of the statistical test depends upon the manner in which the
data is binned. To determine the robustness of our results, we used
both half and double the number of bins to change the resolution of
the sampling. For half the number of bins, we obtain $D = 0.1$
(100\%), $D = 0.1$ (100\%), and $D = 0.2$ (77\%) for Tests 1, 2, and 3
respectively as described above. If we then double the number of bins,
the results are $D = 0.05$ (100\%), $D = 0.0375$ (100\%), and $D =
0.1625$ (22\%) for Tests 1, 2, and 3 respectively. Clearly the results
for Tests 1 and 2 are consistent with previous results and the results
for Test 3 retain their discrepancies though with a variety of values.
As described earlier, Test 3 ($b = 0.8$) is the least relevant of the
results since the mean impact parameter is $b = 0.5$.

\begin{table}
  \begin{center}
    \caption{Minimum eccentricities for selected candidates.}
    \label{ecctab}
    \begin{tabular}{@{}rcccc}
      \hline
      KOI & Period & $t_{\mathrm{kepler}}$ & $\Delta t$ &
      $e_{\mathrm{min}}$ \\
          & (days) & (hours)               & (hours)    & \\
      \hline
        44.01 &  66.47 & 19.74 &  12.2 & 0.74 \\
       211.01 & 372.11 &  4.81 &  10.5 & 0.82 \\
       625.01 &  38.14 &  4.24 &  10.7 & 0.85 \\
       682.01 & 562.14 &  9.49 &  10.8 & 0.64 \\
      1230.01 & 165.72 & 27.26 &  11.8 & 0.34 \\
      1894.01 &   5.29 &  8.80 &  14.4 & 0.75 \\
      2133.01 &   6.25 & 11.26 &  14.1 & 0.67 \\
      2481.01 &  33.85 & 14.95 &  16.8 & 0.64 \\
      \hline
    \end{tabular}
  \end{center}
\end{table}

Finally, we investigate a sample of the outliers with particularly
large deviations from the circular model ($\Delta t >
10$~hours). These candidates are shown in Table \ref{ecctab}. Since
the Kepler data frequently do not contain any secondary eclipse, $e$
and $\omega$ are unknown. We calculate transit duration
$t_{\mathrm{ecc}}$ as a function of $e$ and $\omega$ via Equation
\ref{scaling}. We then produce a grid of $| t_{\mathrm{circ}} -
t_{\mathrm{kepler}} | / |t_{\mathrm{circ}} - t_{\mathrm{ecc}} |$ for
all values of $e$ and $\omega$.  Locations where the grid values are
approximately equal to 1 are possible solutions for which the measured
transit duration in the Kepler candidate catalog is consistent with
certain values of $e$ and $\omega$.

\begin{figure}
  \includegraphics[angle=270,width=8.2cm]{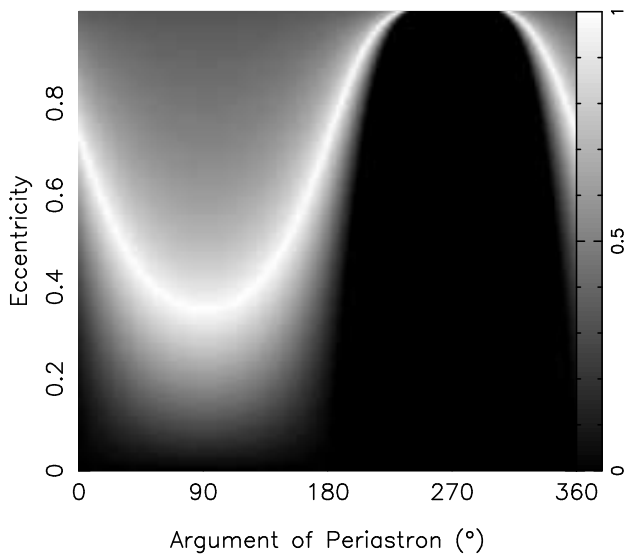}
  \caption{An intensity map for KOI~1230.01 that shows the result of
    dividing $\Delta t$ using $t_{\mathrm{kepler}}$ by $\Delta t$
    using $t_{\mathrm{ecc}}$. Thus, a value of 1 (peak intensity)
    corresponds to the best solution (\S \ref{kepler}).}
  \label{evsw}
\end{figure}

An example of this is shown in Figure \ref{evsw} where we present
results of the above calculations as an intensity map for
KOI~1230.01. In order to be compatible with the Kepler measured
duration, the eccentricity of the planet must be at least 0.34. This
process is repeated for each of the candidates in Table \ref{ecctab}
in which we report the minimum required eccentricity
$e_{\mathrm{min}}$ for each candidate. It is worth noting, however,
that these minimum eccentricities are not singular values but rather
distributions, as can be seen by the gray-scale in Figure
\ref{evsw}. The uncertainties depend highly upon the various random
errors in the measured values of the Kepler candidates catalogue,
including $i$. For example, the stellar radius of KOI~2481.01 would
need to be $\sim 45$\% of the catalogue value in order for it to be in
a circular orbit and the duration discrepancy to be reduced to zero.

Further of interest in Table \ref{ecctab} are the relatively
short-period planets KOI~1894.01 and KOI~2133.01. One normally expects
a transit duration of several hours for period such as these. However,
the values of $t_{\mathrm{kepler}}$ and $\Delta t$ shown in
this table imply a $t_{\mathrm{circ}}$ larger than 20 hours! This does
not appear to make sense until one considers the stellar radius. Note
from Equation \ref{duration} that, for an edge-on orbit and small
$R_p$, the transit duration scales linearly with the size of the
star. For these two candidates, the stellar radii are 8.6 and 9.3
solar radii respectively thus resulting in a large $\Delta t$
and a significant eccentricity required to be consistent with
observations. Note, however, that we have assumed $b=0$. As one
increases the impact parameter, the predicted transit duration will
decrease and thus become closer to its measured value. Results for
individual cases extracted from the global distribution, such as those
in Table \ref{ecctab}, must therefore be treated with caution.

%%%%%%%%%%%%%%%%%%%%%%%%%%%%%%%%%%%%%%%%%%%%%%%%%%%%%%%%%%%%%%%%%%%%

\section{Planet Size Correlation}
\label{correlation}

\begin{figure}
  \includegraphics[angle=270,width=8.2cm]{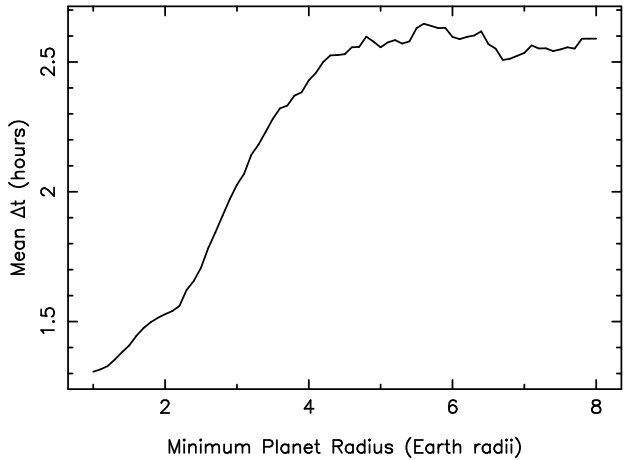}
  \caption{The mean $\Delta t$ for the Kepler candidates as a function
    of minimum planetary radius included in the sample (\S
    \ref{correlation}).}
  \label{mean}
\end{figure}

We mentioned in Section \ref{kepler} that the analysis of the Kepler
objects included only candidates for which $R_p > 8 R_\oplus$. Here we
perform a separate study by repeating the calculations of $\Delta t$
for the Kepler candidates for a range of planetary radii. We allow the
candidate sample to include all radii larger than $1 R_\oplus$ to $8
R_\oplus$ and calculate the mean of the $\Delta t$ distribution in
each case. We show our results in Figure \ref{mean}.

An interpretation of this figure is that the eccentricity distribution
of exoplanets remains relatively flat until we probe below planets the
size of Neptune. At that point the eccentricity distribution of the
orbits becomes rapidly and significantly more circular. This is not
unexpected since we understand from the solar system that the
inefficiency of tidal dissipation (the quality factor $Q$) is much
larger for high-mass than for low-mass planets \citep{gol66},
resulting in shorter tidal circularization time scales for smaller
planets. One aspect of the Kepler candidate sample that may influence
this result is that they are dominated by planets at smaller
semi-major axes since these have much larger transit probabilities. As
indicated by \citet{lis11b} and \citet{lis12}, multi-planet systems
comprise a large proportion of the total Kepler candidate sample and
these systems in particular are less prone to be false-positives. The
findings that planet occurrence increases with decreasing planet mass
\citep[see for example][]{how10} then suggests that smaller planets
find stable architectures in systems with circular orbits and without
large planets in eccentric orbits. This lends credence to two
scenarios: (1) core accretion forming terrestrial planets in circular
orbits, and (2) disk instability and capture scenario explaining the
existence of giant planets in eccentric orbits.

A potential alternative explanation for the dependence of
$\Delta t$ upon $R_{planet}$ is a correlation between planetary
radii and semi-major axes in the Kepler sample due to
completeness. Smaller planets in closer orbits would be more likely to
have had their orbits circularized leading to their domination of the
sample. We do not, however, see such an effect but instead find no
correlation between planet size and period in the Kepler catalog for
the selected range of orbital periods. Thus, we conclude that this
effect is not the cause of the observed radius/eccentricity
correlation.

%%%%%%%%%%%%%%%%%%%%%%%%%%%%%%%%%%%%%%%%%%%%%%%%%%%%%%%%%%%%%%%%%%%%

\section{Discussion}
\label{discussion}

There are various sources of potential systematic noise inherent in
the data used to perform this analysis. For example, we have not taken
the stellar limb-darkening into account when considering transit
durations. However, since we only consider the total transit duration
(first contact to last contact), this will be a negligible effect.

For the Kepler candidates, the primary source of uncertainty arises
from the stellar parameters that are used to derive many of the
planetary candidate parameters. The primary difficulties arise from
the stellar radii whose precision is usually worse than $\sim$10\%,
even when spectra are available. Our assumption is that these
uncertainties are not significantly biased in one direction of the
other and thus only add white noise to the overall statistical
properties. One method to test this assumption is to consider
multi-candidate systems for which a change in the stellar radius will
effect all calculated transit durations in a similar way. For example,
KOI~157 (Kepler-11) has 6 detected planet candidates with measured
transit durations that are shorter than predicted, and with an
estimated host star radius of 1.06~$R_\odot$. Three of those
discrepancies are quite small making them almost consistent with a
circular orbit, as noted by \citet{lis11a}. Attempting to force
circular orbits by reducing the stellar radius slightly makes 2 of the
planets consistent with a circular orbit but leaves the other
durations highly discrepant. A more global test of the assumption is
the application of a range of uniform scaling factors to the stellar
radii to try to produce transit durations consistent with circular
orbits for the bulk of the distribution. These corrections did not
change the distribution shown in Figure \ref{keplercand}, leading to
the conclusion that our calculated $\Delta t$ values are not affected
by systematically incorrect stellar radii.

Considering the uncertainties in the radii of the Kepler host stars,
our limits on the eccentricities of the specific Kepler candidates
discussed in Section \ref{kepler} should be treated in that
context. Indeed, this is why we concentrate our comments on the global
distribution of all the Kepler objects and their parameters, rather
than on individual cases. Furthermore, the Kepler sample is only
complete to $\sim 0.5$~AU with declining completeness beyond this to
$\sim 1.5$~AU. Thus, the number of candidates beyond 0.5~AU is smaller
but still sufficient for a valid comparison to be made.

One aspect of the Kepler candidates that was not taken into account
was the multiplicity of the systems. As suggested at the conclusion of
the previous section, the multiplicity may indeed play a significant
role in stabilizing planets in approximately circular orbits,
particularly for those in the low mass/size regime. This is true of
the radial velocity planets also, some of which are known to lie in
multiple systems of super-Earth mass planets and with relatively
circular orbits such as the system orbiting HD~10180 \citep{lov11}.

In Section \ref{intro} we mentioned the eccentricity bias found by
\citet{she08} due to low-amplitude signals in RV samples. This is a
small but real effect which depends upon the sampling rate and has the
consequence of under-estimating the number of near-circular
orbits. \citet{she08} develop a figure-of-merit and find that only
$\sim 10$\% of the planets considered in their sample are affected by
this bias. The samples studied here are two small to detect such an
effect, but we mention it here as a consideration for future similar
work for which the sample sizes and, more particularly, the period
range explored have grown substantially.

Finally, a minor impact on the stellar radii that should be noted is
the one due to the relation between planet frequency and stellar
metallicity. \citet{joh10} explored the mass-metallicity relationship
for stars that harbor planets and found a positive correlation of
planet frequency with both stellar mass and metallicity, in accordance
with the findings of \citet{fis05}. This positive correlation was also
found empirically for M dwarfs by \citet{ter12}. For a given stellar
mass, a larger metallicity leads to a smaller radius in order to reach
hydrostatic equilibrium. The implication for this study is that many
of the Kepler host stars will have relatively high metallicity leading
to an over-estimated radius. However, this effect is at the level of a
few percent and not expected to interfere with the results of this
study.

%%%%%%%%%%%%%%%%%%%%%%%%%%%%%%%%%%%%%%%%%%%%%%%%%%%%%%%%%%%%%%%%%%%%

\section{Conclusions}
\label{conclusion}

By conducting a transit survey that is sensitive to long enough
periods, it is expected that one will eventually reproduce the
eccentricity distribution found amongst radial velocity planets.  This
has not been possible until very recently, when the Kepler presented a
large sample of long-period planets candidates, providing the
incentive for this study and a similar one by \citet{pla12}. For
individual planets, the eccentricity may be discerned via asymmetry in
the shape of ingress and egress \citep{kip08}. This requires exquisite
photometry and is highly sensitive to $\omega$. We have shown here the
consistency of the Kepler candidates' eccentricity distribution with
their RV planets counterparts. The correlation of eccentricity with
planet size is also an expected result based upon the discoveries of
small planets in multiple systems and indicates that there is an
empirical approach from which to both reverse-engineer formation
scenarios and predict future stability patterns.

%%%%%%%%%%%%%%%%%%%%%%%%%%%%%%%%%%%%%%%%%%%%%%%%%%%%%%%%%%%%%%%%%%%%

\section*{Acknowledgements}

The authors would like to thank Rory Barnes, Eric Ford, Andrew Howard,
and Suvrath Mahadevan for several useful discussions. We would also
like to thank the anonymous referee, whose comments greatly improved
the quality of the paper. This research has made use of the Exoplanet
Orbit Database and the Exoplanet Data Explorer at exoplanets.org. This
research has also made use of the NASA Exoplanet Archive, which is
operated by the California Institute of Technology, under contract
with the National Aeronautics and Space Administration under the
Exoplanet Exploration Program.

%%%%%%%%%%%%%%%%%%%%%%%%%%%%%%%%%%%%%%%%%%%%%%%%%%%%%%%%%%%%%%%%%%%%

\end{document}